\documentclass[pra,showpacs,preprintnumbers,amsmath,amssymb]{revtex4}
\usepackage{amsmath}
\usepackage{epsf}
\usepackage{graphicx}

\newcommand{\ket}[1]{\mathop{\left| #1 \right\rangle}\nolimits}

\begin{document}
\title{Triplet--like correlation symmetry of continuous variable entangled states}
\author{Gerd~Leuchs, Ruifang~Dong, and Denis~Sych}
\affiliation{Max Planck Institut f\"ur die Physik des Lichts, G\"unther--Scharowsky--Strasse 1 / Bau24,
D-91058 Erlangen, Germany}
\affiliation{Institut f\"ur Optik, Information und Photonik, Universit\"at Erlangen--N\"urnberg, Staudtstrasse 7 / B2, 91058 Erlangen, Germany}
\begin{abstract}
We report on a continuous variable analogue of the triplet two--qubit Bell states. We theoretically and experimentally demonstrate a remarkable similarity of two--mode continuous variable entangled states with triplet Bell states with respect to their correlation patterns. Borrowing from the two qubit language, we call these correlations triplet--like. 
\end{abstract}
\pacs{03.65 Ud, 03.65 Vf, 03.67 Mn}
\maketitle


\section{Introduction}

Entanglement has enjoyed a special role in physics ever since the famous discussion between Einstein and Bohr about their Gedankenexperiments~\cite{Einstein:35,Bohr:35}: a pair of sub systems was postulated in which the ability to infer the values of two conjugate observables of the second sub system based on observations of the first sub system is less uncertain than the uncertainty allowed by quantum mechanics in the case of a single isolated subsystem. Such counter intuitive states exhibit correlations between two sub-systems of a very non-classical nature and in turn lead to fundamentally new interactions and applications in the field of quantum infomation~\cite{BEZ:00,Bennett:00Nat}. Launched by this essential role in quantum information processing, a great number of experiments have investigated the production of entangled states of photons, both in discrete variable ({\em DV}) regime and in continuous variable ({\em CV}) regime. 

For a long time DV and CV states were investigated separately from each other. Recently, the developments of quantum information applications such as quantum cryptography, quantum computations and others, exploit both ``worlds''~--- discrete and continuous, as far as both have some advantages \cite{Peres:96,Horodecky:96,Duan:00,Simon:00,Braunstein:05,Scarani:08}. This stimulates further theoretical and experimental research aimed to unify the ``worlds''.

We found an interesting similarity between DV and CV states, based on the rotational invariance of the bipartite states. It is well known that the symmetry pattern of rotational invariance of DV entangled states plays a very special role in the theoretical analysis of entanglement \cite{Werner:89,Horodecky:99,Vollbrecht:01,Audenaert:01,Sych:09}. We show a continuous variable analogy of the triplet Bell states in the sense of their symmetry. We theoretically demonstrate a similarity between the correlation pattern of the triplet Bell states and a two--mode continuous variable entangled state. We investigate experimentally a two--mode CV polarization entangled source and the correlations between the two modes along different detection directions. The measured correlations show the same symmetry with those of discrete triplet Bell states.

\section{Correlation properties of two--mode states}
\label{Sec:CVent}
To describe a mode of the electromagnetic field we use the creation and annihilation operators ($\hat a^{\dag}$ and $\hat a$) and a pair of canonical conjugate quadratures ($\hat X$ and $\hat P$). We follow a notation where $\hat a=\hat X+i\hat P$. Let us consider two modes $A$ and $B$ mixed on a $50/50$ beamsplitter with a relative phase of $\pi/2$. The resulting output modes (after the beamsplitter) are labeled as $C$ and $D$, respectively. The beamsplitter transformation is $\hat a_{C,D}=\left(\hat a_{A}\pm i \hat a_{B}\right)/\sqrt{2}$, thus the output conjugate quadratures are
\begin{equation}\label{outputmodes}
 \begin{array}{c}
 \hat X_{C,D}=\left(\hat X_{A}\mp \hat P_{B}\right)/\sqrt{2}\\
 \hat P_{C,D}=\left(\hat P_{A}\pm \hat X_{B}\right)/\sqrt{2}.
 \end{array}
\end{equation}
Now we consider a case when both input modes are squeezed in $\hat X$ direction in phase space: $\hat X_{A}^{(r)}=e^{-r}\hat X_{A}^{(0)}\rightarrow 0$ and $\hat X_{B}^{(r)}=e^{-r}\hat X_{B}^{(0)}\rightarrow 0$, where $r$ is a parameter of squeezing. Then the output modes (\ref{outputmodes}) becomes individually noisy but conditionally quiet:
\begin{equation}\label{CVentpm}
 \begin{array}{c}
 \hat X_{C}+\hat X_{D}=\sqrt{2} \hat X_{A}\rightarrow 0\\
 \hat P_{C}-\hat P_{D}=\sqrt{2} \hat X_{B}\rightarrow 0.
 \end{array}
\end{equation}

Let us define a quadrature with an arbitrary phase $\phi$ as $\hat X(\phi)=\cos(\phi)\hat X+\sin(\phi)\hat P$. Then the correlations (\ref{CVentpm}) can be shortly rewritten as 
\begin{equation}\label{CVentshort}
\hat X_{C}(\phi)+\hat X_{D}(-\phi)=\\\sqrt{2}\cos(\phi)\hat X_{A}+\sqrt{2}\sin(\phi)\hat X_{B}\rightarrow 0.
\end{equation}

The physical essence of this expression is a clear correlation symmetry of two entangled output modes $A$ and $B$. A quadrature of the mode $A$ with a phase $\phi$ is correlated with a ``mirror'' quadrature of the mode $B$ with a phase $-\phi$. In other words, any measurement result of a quadrature $\hat X_{C}(\phi)$ is opposite (in the limit $r\rightarrow\infty$) to the measurement result of a ``mirror'' quadrature $\hat X_{D}(-\phi)$. As we show below, this type of correlations is similar to the triplet Bell states, thus we call it ``triplet--like''.

\section{Correlation properties of two--qubit states}

In the DV case, the basis of maximally entangled states consists of four Bell states~---the fully isotropic singlet state and three triplet states
\begin{equation}
\Phi^{\pm}=(\ket{0}_A\ket{0}_B\pm\ket{1}_A\ket{1}_B)/\sqrt{2},\quad \Psi^{\pm}=(\ket{0}_A\ket{1}_B\pm\ket{1}_A\ket{0}_B)/\sqrt{2},
\label{BellStates}
\end{equation}
corresponding to the total spin $S=0$ and the $S=1$ manifold. The property of rotational invariance of the singlet Bell state $\Psi^-$ is widely known and used in various experiments with qubits. Other Bell states, so called triplet states $\Psi^+, \Phi^\pm$, don't have this property. Instead, they have a different symmetry which is also used in various quantum information tasks such as hidden--variable models, separability criterion, entanglement distillation and characterization  \cite{Werner:89,Horodecky:99,Vollbrecht:01,Audenaert:01}. 

As we show in Appendix~\ref{AppBellSymm}, all triplet states follow certain $U\otimes U^{\ast}$ invariances, which means, that if one applies a unitary operation $U$ on the first subsystem and a unitary operation $U^{\ast}$ on the second subsystem, the joint state remains unchanged. Consequently, when measuring in a basis obtained by a $U\otimes U^{\ast}$ transformation of the standard computational basis, the observed correlations between local measurements on the sub systems are unchanged. As we show in Appendix~\ref{AppBellSymm}, the relative symmetry between two perfectly correlated measurements is either central symmetry (for the singlet state) or mirror symmetry (for the triplet states). The latter case of mirror symmetry appears to be very similar to the ``triplet--like'' symmetry of continuous variable states, which is described in the previous section.

\section{Description of the experiment}
\noindent In the experiment we investigate the correlation pattern underlying the two-mode CV entangled states and demonstrate its triplet-like nature. 
Instead of the usual quadrature variables in single mode phase space, we perform the experiment in two mode polarization space described by the Stokes variables $S_1, S_2, S_3$. These polarization Stokes variables have the distinct experimental advantage over the quadrature variables~\cite{Silberhorn:01,Josse:03, Heersink:03} that the characterization of all relevant parameters of the polarization variables is possible by linear optical manipulations and direct detection~\cite{Korolkova:02}. Therefore, a highly efficient continuous variable polarization entangled source is used for the experimental demonstration~\cite{Dong:07}.

\begin{figure*}[ht]
  \centerline{
    \includegraphics[width = 11cm]{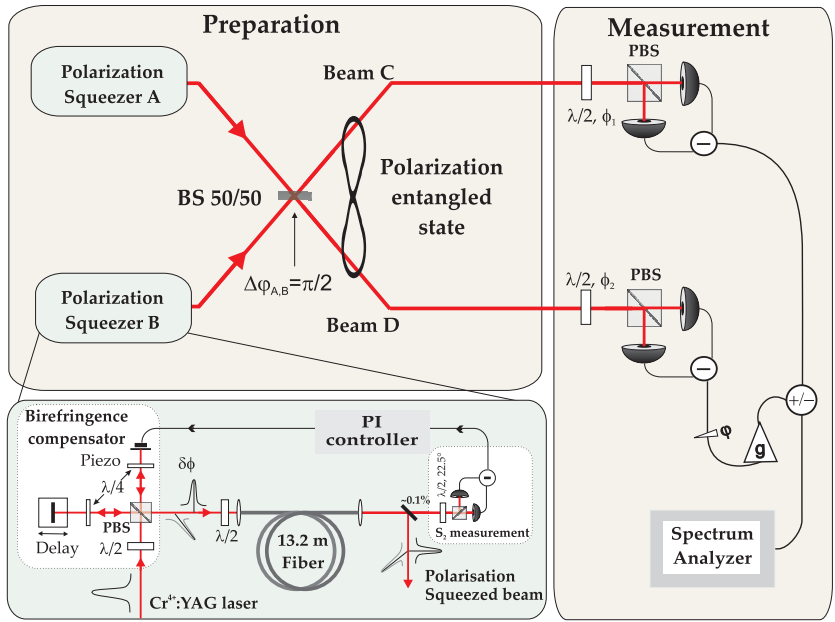}
	}
	\caption{Schematics of the experimental setup for the preparation and measurement of the polarization entanglement. Experimental setup for efficient polarization squeezing generation is depicted in the inset. Two polarization squeezed beams interfere at a 50:50 beam splitter with a relative phase of $\Delta \varphi_\mathrm{A,B}=\pi/2$. In the two output ports, A~and~B, the dark plane Stokes parameters $\hat{S}_{A}(\phi_1)$ and $\hat{S}_{B}(\phi_2)$ are measured, and the photo-currents are added or subtracted to check for correlations. A variable gain g and phase shift $\varphi$ is introduced in the cables to minimise the variances. PBS, polarizing beam splitter; $(\lambda /2)$, half-wave plates; $(\lambda /4)$, quarter-wave plates.}
\label{setup}
\end{figure*}

The experimental setup is shown in Fig.~\ref{setup}. The preparation of the highly entangled polarization state is accomplished by two steps: generating two independent polarization squeezed beams A and B with the same noise properties, and producing polarization entanglement by interference of the two beams on a 50/50 beam splitter (BS). In the inset of Fig.~\ref{setup}, the setup for the generation of polarization squeezed beam is shown: A home-made Cr$^{4+}$:YAG laser with a central wavelength of 1497~nm was used to produced soliton shaped pulses at a repetition rate of 163~MHz with a duration of 140~fs. These pulses were measured to be shot noise limited at our measurement frequency (17.5MHz) and thus can be assumed to be coherent. We exploit the method based on single pass of two orthogonally polarised light pulses through a birefringent fibre (13.2~meters of 3M FS-PM-7811, mode field diameter 5.7~$\mu$m, beat length 1.67~mm)~\cite{Heersink:05, Dong:08}, two quadrature squeezed states were then independently generated. 

The two light pulses were adjusted to have the same optical powers after the fiber, the generated squeezed states thus have approximately the same quadrature noise property (the squeezed quadrature is skewed by angle $\theta_{sq}$ from the amplitude quadrature). To generate the polarization squeezing, the two emerging pulses were overlapped with a $\pi/2$ relative phase shift. This was accomplished by using an active phase lock in the pre-compensation of the fibre birefringence which introduced a $\delta\phi$ relative shift between the two polarization eigenmodes of the fibre (Fig.~\ref{setup}). This resulted in a circularly polarised beam at the fibre output, mathematically described by $\langle \hat{S}_3 \rangle \neq 0$ and $\langle \hat{S}_1 \rangle = \langle \hat{S}_2 \rangle = 0$. The conjugate polarization operators, which can exhibit polarization squeezing, are then found in the plane given by $\hat{S}_1-\hat{S}_2$, referred to as the "dark plane". We also define a quadrature $S(\theta)=\cos(\theta)S_1+\sin(\theta)S_2$. Our polarization squeezing is derived from Kerr squeezed states in which the squeezed quadrature is skewed by $\theta_\mathrm{sq}$ from the amplitude direction. Thus the squeezed Stokes operator is given by $\hat{S}(\theta_\mathrm{sq})$, and the orthogonal, anti-squeezed Stokes operator is $\hat{S}(\theta_\mathrm{sq}+\pi/2)$, which will be written as $\hat{S}(\theta_\mathrm{asq})$ for convenience~\cite{Heersink:05, Korolkova:02}. These operators both have zero mean values, and they both commute with the bright $\hat{S}_3$ component of the optical field. Therefore, such polarization squeezing is mathematically equivalent to vacuum squeezing, furthermore we can have the correspondance between $\hat{S}(\theta_\mathrm{sq})$ and $\hat{S}(\theta_\mathrm{sq}+\pi/2)$ on the one hand and the two single mode conjugate variables $\hat{X}$ and $\hat{P}$. 

Two such polarization squeezed beams are then simultaneously generated and mixed on a 50:50~beam splitter (Fig.~\ref{setup}). The two resulting intense beams, labeled C and D, are set via a phase lock to have equal intensity, i.e. the two inputs are set to have a $\pi/2$ relative phase shift. Thus the two outputs of the beam splitter become entangled and they are also circularly polarised. The correlation pattern that the polarization entanglement follows between the two modes can be easily achieved according to correspondence between the quadrature variables and the polarization variables, which is expressed by $\hat{S}_C(\theta_\mathrm{sq}+\phi_1) + \hat{S}_D(\theta_\mathrm{sq}+\phi_2)\rightarrow0$ and $\hat{S}_C(\theta_\mathrm{asq}+\phi_1) - \hat{S}_D(\theta_\mathrm{asq}+\phi_2)\rightarrow0$ with $\phi_1=-\phi_2$. These beams are then measured independently in two Stokes measurement apparatuses. Each one is composed of only a half-wave plate ($\lambda/2$) followed by a polarising beam splitter (PBS). Rotation of the half--wave plate allows for the observation of arbitrary Stokes parameters in the dark $\hat{S}_1-\hat{S}_2$ plane, and, therefore, allows for the direct observation of the triplet-like correlation pattern underlying the CV polarization entanglement. In each Stokes measurement the outputs of the PBS are detected by identical pairs of balanced photo--detectors based on custom made pin photo--diodes (98\% quantum efficiency at DC and AC). The detection frequency of 17.5~MHz was chosen to avoid low frequency technical noise as well as the 163~MHz laser repetition rate, although in principle any frequency up to several THz is possible~\cite{Spalter:97}. The four detected AC photocurrents are passively pairwise subtracted. The resulting subtracted pairs are added and monitored on a spectrum analyser (HP 8590E, 300~kHz resolution bandwidth, 30~Hz video bandwidth).

\section{Results}
\noindent Based on the setup as depicted in Fig.~\ref{setup}, we implemented the polarization measurements. The results of the characterisation of a two-mode CV polarization entanglemed source are presented in the following. In the results, the variances $\Delta^2_\mathrm{norm}(\cdot)$ are normalized to the respective mean values of the $\hat S_3$ parameter corresponding to the shot noise reference. As a first step, the polarization squeezing of the two input modes A and B was measured. In order to characterise the initial squeezing of one input mode, we blocked the other input mode and measured the polarization squeezing of the output modes C and D. From the observed level of squeezing, we can infer the amount of squeezing in the input modes. Polarization squeezing of -4.6$\pm$0.3~dB was observed for the $\hat{S}_\mathrm{A}(\theta_\mathrm{sq})$ parameter of source A. Its canonic conjugate, $\hat{S}_\mathrm{A}(\theta_\mathrm{asq})$, was anti--squeezed by +22.3$\pm$0.3~dB. The second beam exhibited similar squeezing levels of -4.5$\pm$0.3~dB in $\hat{S}_\mathrm{B}(\theta_\mathrm{sq})$ and of +22.2$\pm$0.3~dB in $\hat{S}_\mathrm{B}(\theta_\mathrm{asq})$. These noise traces as well as those for the polarization entanglement are corrected for electronic noise. The individual squeezed beams A and B exhibited a total optical power of 9.4~mW, corresponding to an energy of 61~pJ per pulse. The squeezing angle $\theta_\mathrm{sq}$ was $4^\circ$.

\begin{figure*}[ht]
\begin{center}
\includegraphics[width=14cm]{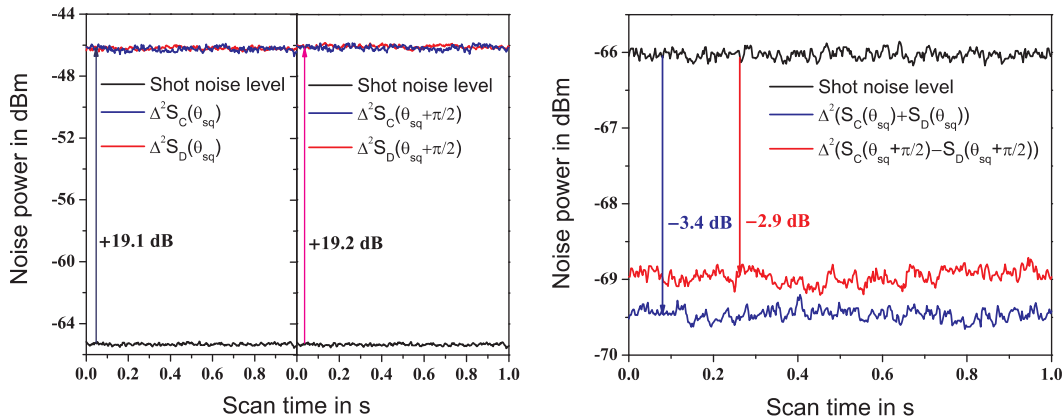}
\end{center}
\caption{Measurement of the noise of the entangled beam pair along $\hat S(\theta_\mathrm{sq})$ and $\hat S(\theta_\mathrm{sq}+\pi/2)$. The noise of the individual beams $\hat S_\mathrm{C,D}(\theta_\mathrm{sq})$ and $\hat S_\mathrm{C,D}(\theta_\mathrm{sq}+\pi/2)$ is plotted on the left side, the correlations $\Delta^2(\hat{S}_\mathrm{C}(\theta_\mathrm{sq}) + \hat{S}_\mathrm{D}(\theta_\mathrm{sq}))$ and $\Delta^2(\hat{S}_\mathrm{C}(\theta_\mathrm{sq}+\pi/2) - \hat{S}_\mathrm{D}(\theta_\mathrm{sq}+\pi/2))$ are plotted on the right side. Note the difference in the level of correlations of the two signals, which is a consequence of an residual asymmetry of the splitting ratio of the entangling beam splitter together with the high level of excess noise. The measurement frequency was $17.5$~MHz, the resolution bandwidth was $300$~kHz and the resolution bandwidth was $30$~Hz.}\label{polent_sq}
\end{figure*}

To generate the maximum polarization entanglement(Fig.~\ref{setup}), the interference visibility between the squeezed input modes A and B was optimized achieving $\geq$98\%. Then we investigate the variances of the conjugate Stokes operators $\hat S(\theta_\mathrm{sq})$ and $\hat S(\theta_\mathrm{sq}+\pi/2)$ of the individual output modes C and D, as plotted in Fig.~\ref{polent_sq}. Each individual mode is seen to exhibit a large excess independent of the angle $\theta$ (around 19 dB have been measured). The sum and difference signals of these two modes are $\Delta^2_{norm}(\hat{S}_\mathrm{C}(\theta_\mathrm{sq})+\hat{S}_\mathrm{D}(\theta_\mathrm{sq}))=-3.4 \pm 0.3$~dB or 0.46$\pm$0.03 and $\Delta^2_{norm}(\hat{S}_\mathrm{C}(\theta_\mathrm{sq}+\pi/2)-\hat{S}_\mathrm{D}(\theta_\mathrm{sq}+\pi/2))=-2.9 \pm 0.3$~dB or 0.51$\pm$0.03, respectively. 

\begin{figure*}[htb]
\begin{center}
\includegraphics[width=14cm]{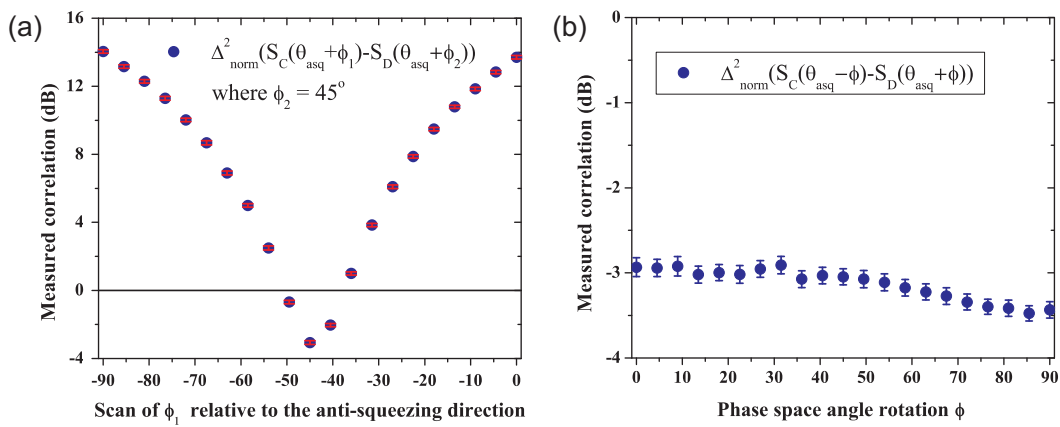}
\end{center}
\caption{Measurement of the correlated variance of the entangled beam pair between the Stokes parameters $\hat S_C(\theta_\mathrm{asq}+\phi_1)$ and $\hat S_D(\theta_\mathrm{asq}+\phi_2)$ for the two cases: (a) the phase rotation angle $\phi_2$ was set to be $45^o$ skewed from the anti-squeezing qudrature, while the $\phi_1$ on beam C side was scanned from 0 to $-90^o$ relative to the anti-squeezing quadrature; (b) both $\phi_1$ and $\phi_2$ were rotated with the same angle but opposite directions for 90 degrees, that is, $-\phi_1=\phi_2=\phi\in(0,90^o)$. }\label{polent_anglerotation}
\end{figure*}

In the next step, we set both $\lambda/2$ wave-plates in the two Stokes measurement setups along the anti-squeezing directions as the reference for the further rotations, and investigated the correlations between the Stokes parameters $\hat S_C(\theta_\mathrm{asq}+\phi_1)$ and $\hat S_D(\theta_\mathrm{asq}+\phi_2)$. To check the correlation dependence on the relevant rotation angles $\phi_{1,2}$, we first rotated the $\lambda/2$ wave-plate in mode D clockwise for $22.25^{o}$, that is, the Stokes parameter observed in beam D side was skewed by $\phi_2=\pi/4$ from the anti-squeezing direction. Then we implemented the measurements of correlation between C and D by scanning $\phi_1$ of the beam C in the counterclockwise direction of from 0 to $-\pi/2 (-90^o)$ with the step of $5^o$. The results are plotted in Fig.~\ref{polent_anglerotation}(a), and we see from the results that, the best correlaton is achieved when $\phi_1=-\phi_2=-45^o$, which corresponds to a measured correlation of $\Delta^2(\hat S_\mathrm{C}(\theta_\mathrm{asq}-45^o)-\hat S_\mathrm{D}(\theta_\mathrm{asq}+45^o)=-3.1\pm0.3$~dB. In addition, we further verified the triplet-like correlation underlying modes C and D by scanning $\phi_1$ and $\phi_2$ in opposite directions with the same rotation step width, i.e., $-\phi_1=\phi_2=\phi$. The results for the observed correlation as a function of rotation angle $\phi$ from the anti-squeezing direction in the phase space are also plotted in Fig.~\ref{polent_anglerotation}(b). Therefore, the correlations between the Stokes parameters $\hat S_C(\theta_\mathrm{asq}-\phi)$ and $\hat S_D(\theta_\mathrm{asq}+\phi)$ were measured to be highly non-classical when $\phi$ was varied from 0 to $\pi/2$. The variation of the observed correlations was attributed to the slight asymmetry of the 50/50 beam splitter (the actual transmittivity was measured as 0.49)~\cite{Dong:07} and the imperfect interference ($\geq$98\% visibility) in the setup for producing the entanglement.  

\section{Summary}
In summary, we found a remarkable similarity between two--mode continuous variable entangled states and the two--qubit triplet Bell states. We demonstrate that even a mixed two--mode continuous variable entangled state with large excess noise in one quadrature shows this similarity. Furthermore, it's the first time that the triplet--like correlation pattern of continuous variable states is discussed and demonstrated experimentally.

\section{Acknowledgement}

The authors thank Jens Eisert, Norbert L\"utkenhaus and Luis Sanchez--Soto for valuable discussions. DS acknowledges Alexander von Humboldt Foundation for a stipend.

\appendix
\section{Symmetry properties of two--qubit Bell states}
\label{AppBellSymm}

\begin{figure}[!ht]
\centerline{\includegraphics[width = 4cm]{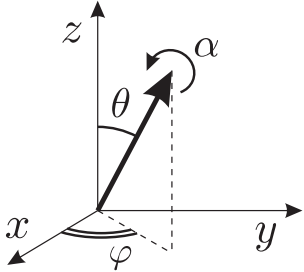}}
\caption{Geometric picture of the general rotation~(\ref{Ugen}) on the Bloch sphere.}
\label{Ugenpic}
\end{figure}

The general rotation $U$ on the Bloch sphere can be parametrized by three angles $\{\theta,\varphi,\alpha\}$ as
\begin{equation}
U(\theta,\varphi,\alpha)=\left(
\begin{array}{cc}
\cos\frac{\alpha}{2}-\cos\theta\sin\frac{\alpha}{2} & (-i \sin\theta\cos\varphi-\sin\theta\sin\varphi)\sin\frac{\alpha}{2}\\
(-i\sin\theta\cos\varphi+\sin\theta\sin\varphi)\sin\frac{\alpha}{2} & \cos\frac{\alpha}{2}+i\cos\theta\sin\frac{\alpha}{2})
\end{array}
\right).
\label{Ugen}
\end{equation}
Geometrically, two angles $\{\theta,\varphi\}$ define an axis orientation, and the angle $\alpha$ defines rotation around this axis (see Fig.~\ref{Ugenpic}).

Applying the transformation $U\otimes U$ to the singlet state $\Psi^-$ one can check its invariance. The triplet states don't hold this property. Instead, we look at the lower symmetry with respect to the rotations around $x-,y-$ and $z-$axis by an angle $\alpha$. Specifying $x$, $y$ or $z$ as the rotation axis, $U(\theta,\varphi,\alpha)$ (Eq. \ref{Ugen}) turns into:
\begin{equation}
U_x=\left(
\begin{array}{cc}
\cos\frac{\alpha}{2} & -i\sin\frac{\alpha}{2}\\
-i\sin\frac{\alpha}{2} & \cos\frac{\alpha}{2}
\end{array}
\right), \quad
U_y=\left(
\begin{array}{cc}
\cos\frac{\alpha}{2} & -\sin\frac{\alpha}{2}\\
\sin\frac{\alpha}{2} & \cos\frac{\alpha}{2}
\end{array}
\right), \quad
U_z=\left(
\begin{array}{cc}
e^{-i\frac{\alpha}{2}} & 0\\
0 & e^{i\frac{\alpha}{2}}
\end{array}
\right).
\label{Uxyz}
\end{equation}

By direct calculations of the transformations $U_i\otimes U_i$ and $U_i\otimes U_i^{\ast}$ we observe that the Bell states demonstrate a particular type of invariance, which is summarized in Table~\ref{SymTab}.

\begin{table}[!ht]
\caption{Symmetric invariances of the Bell states}
\label{SymTab}
\begin{tabular}{|c|c|c|c|c|c|c|}
\hline
& $U_x\otimes U_x$ & $U_y\otimes U_y$ & $U_z\otimes U_z$ & $U_x\otimes U_x^{\ast}$ & $U_y\otimes U_y^{\ast}$ & $U_z\otimes U_z^{\ast}$\\
\hline
$\Psi^-$ & yes & yes & yes & no & no & no \\
\hline
$\Psi^+$ & no & no & yes & yes & yes & no \\
\hline
$\Phi^-$ & yes & no & no & no & yes & yes \\
\hline
$\Phi^+$ & no & yes & no & yes & no & yes \\
\hline
\end{tabular}
\end{table}
We can see, that each of the Bell state is invariant under three rotations out of six. All triplet states are similar to each other in a sense that they are invariant under two $U\otimes U^{\ast}$ rotations out of three, and additionally have one complementary $U\otimes U$ symmetry. The transformations $U^{\ast}$ which are conjugate to (\ref{Uxyz}) can be rewritten as a reverse rotation by the same angle: $U^{\ast}(\alpha)=U(-\alpha)$. Here we can clearly see the similarity of the triplet Bell states and CV entangled states described in Section~\ref{Sec:CVent}.

The general rotation~(\ref{Ugen}) is parametrized by three angles $\{\theta,\varphi,\alpha\}$, however a point on the sphere can be specified by only two of them, e.g. $\{\theta,\varphi\}$. A particular transformation from one point to another point can be achieved by two appropriate consecutive transformations~(\ref{Uxyz}), i.e. can be specified by two parameters $\{\theta,\varphi\}$. Respectively, a complimentary operation $U^{\ast}$ can be specified by another pair of $\{\theta\prime,\varphi\prime\}$. 

\begin{figure}[!ht]
\centerline{\includegraphics[width = 5cm]{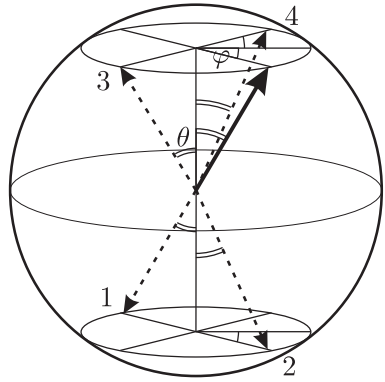}}
\caption{Geometric picture of the correlation properties of the Bell states. The thick arrow represents a measurement result of the first subsystem on the Bloch sphere. The dashed arrows 1, 2, 3 and 4 represent perfectly correlated results of a measurement of the second subsystem for the states $\Psi^-$, $\Psi^+$, $\Phi^-$, and $\Phi^+$ respectively.}
\label{BellBloch}
\end{figure}

Symmetry properties of the Bell states presented in Table~\ref{SymTab} can be visualized geometrically on the Bloch sphere (Fig.~\ref{BellBloch}).
From this figure we can explicitly see that the symmetry properties of the Bell states are of two distinctive types: singlet and triplet. The singlet state $\Psi^-$ (the arrow 1 in Fig.~\ref{BellBloch}) follows inversion symmetry, i.e. reflection at the centre, and the two correlated states are oriented in opposite directions. The triplet states (the arrows 2, 3 and 4 in Fig.~\ref{BellBloch}) follow mirror reflection symmetry, i.e. two correlated states are mirror reflected in a certain plane ($xy$, $yz$, or $xz$). 


\bibliography{referencesall}

\end{document}